\documentclass[journal]{IEEEtran}

\usepackage[noadjust]{cite}
\usepackage[dvips]{graphicx}
\usepackage[cmex10]{amsmath}
\usepackage{amsfonts}
\usepackage{amsthm}

\usepackage{tikz}
\usetikzlibrary{arrows, decorations.pathmorphing, backgrounds, positioning, fit}

\newtheorem{mydef}{Definition}
\newtheorem{mylem}{Lemma}
\newtheorem{mythe}{Theorem}
\newtheorem{myrem}{Remark}

\newtheorem{mycol}{Corollary}

\newtheorem{myass}{Assumption}

\newcommand{\bseq}{\begin{subequations}}
\newcommand{\eseq}{\end{subequations}}
\newcommand{\bmat}{\begin{bmatrix}}
\newcommand{\emat}{\end{bmatrix}}
\newcommand{\beq}{\begin{equation}}
\newcommand{\eeq}{\end{equation}}
\newcommand{\beqs}{\begin{equation*}}
\newcommand{\eeqs}{\end{equation*}}
\newcommand{\bali}{\begin{aligned}}
\newcommand{\eali}{\end{aligned}}
\newcommand{\RR}{\mathbb{R}}

\newcommand{\barr}{\begin{array}}
\newcommand{\earr}{\end{array}}

\newcommand{\ha}{\hat{a}}
\newcommand{\hb}{\hat{b}}
\newcommand{\hc}{\hat{c}}
\newcommand{\hk}{\hat{k}}
\newcommand{\din}{\textrm{dim}}
\newcommand{\diag}{\textrm{diag}}

\newcommand{\al}{\alpha}
\newcommand{\ga}{\gamma}

\begin{document}
\title{Network Identifiability from Intrinsic Noise} 

\author{David~Hayden, Ye~Yuan and~Jorge~Gon\c{c}alves%
\thanks{D. Hayden and Y. Yuan are with Department of Engineering, University of Cambridge, CB2 1PZ, UK, email: {dph34, yy311@cam.ac.uk}.}%
\thanks{J. Gon\c{c}alves is with Department of Engineering, University of Cambridge, CB2 1PZ, UK and University of Luxembourg, Facult\'{e} des Sciences, de la Technologie et de la Communication, 7 Avenue des Hauts Fourneaux, L-4362 BELVAL, Luxembourg, email: {jmg77@cam.ac.uk}.}%
\thanks{This work was supported by the Engineering and Physical Sciences Research Council under Grants EP/G066477/1 and EP/I03210X/1.}}
\maketitle

\begin{abstract}
This paper considers the problem of inferring an unknown network of dynamical systems driven by unknown, \emph{intrinsic}, noise inputs. Equivalently we seek to identify direct causal dependencies among manifest variables only from observations of these variables. For linear, time-invariant systems of minimal order, we characterise under what conditions this problem is well posed. We first show that if the transfer matrix from the inputs to manifest states is minimum phase, this problem has a unique solution irrespective of the network topology. This is equivalent to there being only one valid spectral factor (up to a choice of signs of the inputs) of the output spectral density.

If the assumption of phase-minimality is relaxed, we show that the problem is characterised by a single Algebraic Riccati Equation (ARE), of dimension determined by the number of latent states. The number of solutions to this ARE is an upper bound on the number of solutions for the network. We give necessary and sufficient conditions for any two dynamical networks to have equal output spectral density, which can be used to construct all equivalent networks. Extensive simulations quantify the number of solutions for a range of problem sizes. For a slightly simpler case, we also provide an algorithm to construct all equivalent networks from the output spectral density.
 \end{abstract}

\section{Introduction}

\IEEEPARstart{M}{any} phenomena are naturally described as networks of interconnected dynamical systems and the identification of such networks remains a challenging problem, as evidenced by the diverse literature on the subject. In biological applications in particular, experiments are expensive to conduct and one may simply be faced with the outputs of an existing network driven by its own intrinsic variation. Noise is endemic in biological networks and its sources are numerous \cite{raser05}; making use of this natural variation as a non-invasive means of identification is an appealing prospect, for example in gene regulatory networks \cite{dunlop08}. We now give a brief overview of relevant work, focusing on the case where the network is driven by stochastic, rather than deterministic inputs.

An active problem in spectral graph theory is whether the topology of a graph can be uniquely determined from the spectrum of, for example, its adjacency matrix. There are simple examples of non-isomorphic graphs for which this is not possible and classes of graph for which it is (see \cite{vanDam03}); however the approach does not consider dynamics of the graph (other than that of the adjacency matrix). A related problem is that considered in the causality literature \cite{pearl} of determining a graph of causal interactions between events from their statistical dependencies. Again, there are classes of graph for which this is possible and again the dynamics of the system are not considered. Probabilistic methods, such as \cite{PN}, seek to identify a network in which each state is considered conditionally independent of its non-descendants, given its parent states. An heuristic search algorithm is then used to select an appropriate set of parent states, and hence obtain a graph of dependencies.

Granger \cite{granger} considered the problem of determining causality between a pair of states interacting via Linear, Time-Invariant (LTI) systems. An autoregressive model is estimated for the first state, then if the inclusion of the second state into the model significantly improves its prediction, the second state is said to have a causal influence on the first. The idea can be extended to networks of greater than two states by considering partial cross spectra, but it is difficult to guarantee that this problem is well posed -- there could be multiple networks that explain the data equally well. Two crucial issues are the choice of model order and the combinatorial problem of considering all partial cross spectra \cite{Atuk, Michael}.

Current approaches to network reconstruction that offer guarantees about the uniqueness of the solution require either that assumptions about the topology be made or that the system dynamics are known. For example, in \cite{ren} the undirected graph of a network of coupled oscillators can be found if the system dynamics and noise variance are known. In \cite{preci}, networks of known, identical subsystems are considered, which can be identified using an exhaustive grounding procedure similar to that in \cite{nabi}. A solution is presented in \cite{mati_cdc12} for identifying the undirected structure for a restricted class of polytree networks; and in \cite{mati_tac} for \lq\lq self-kin\rq\rq \ networks. In contrast, for networks of general, but known topology, the problem of estimating the dynamics is posed as a closed-loop system identification problem in \cite{VdH_autom}.

We focus on LTI systems with both unknown and unrestricted topology and unknown dynamics and consider the problem from a system identification perspective. The origin of this problem is arguably the paper by Bellman and Astrom \cite{astrom} in which the concept of \emph{structural identifiability} is introduced. A model is identifiable if its parameters can be uniquely determined given a sufficient amount of data, which is a challenging problem for multivariable systems \cite{ljung}. Previous work has characterised the identifiability of a network of LTI systems in the deterministic case where targeted inputs may be applied \cite{TAC08}. The network was modelled as a single transfer matrix representing both its topology and dynamics; the network reconstruction problem is then well posed if this transfer matrix is identifiable.

The purpose of this paper is to assess the identifiability of networks with unknown, stochastic inputs. The identifiability of state-space models in this setting is considered in \cite{keith74} based on the spectral factorization results of \cite{anderson69}, in which all realizations of a particular spectral density are characterized. We present novel results on the relationship between an LTI network and its state-space realizations and use these to characterise all solutions to the network reconstruction problem.

Our contributions are threefold: first, for networks with closed-loop transfer matrices that are minimum phase, we prove that the network reconstruction problem is well posed -- the network can be uniquely determined from its output spectral density; second, in the general case, we provide an algebraic characterization of all networks with equal output spectral density, in which every network corresponds to a distinct solution to an Algebraic Riccati Equation; and third, for a slightly simpler case, we provide an algorithm to construct all such solutions from the spectral density.

Section \ref{sec:pre} provides necessary background information on spectral factorization, structure in LTI systems and the network reconstruction problem. The main results are then presented in Section \ref{sec:main}, followed by a detailed example and numerical simulations in Section \ref{sec:sim}. One further case is considered in Section \ref{sec:full}, in which noise is in addition applied to the latent states. Conclusions are drawn in Section \ref{sec:conc} and additional proofs are included in the Appendix.

\subsection*{Notation}
Denote by $A(i,j)$, $A(i,:)$ and $A(:,j)$ element $(i,j)$, row $i$ and column $j$ respectively of matrix $A$. Denote by $A^T$ the transpose of $A$ and by $A^*$ the conjugate transpose. We use $I$ and $0$ to denote the identity and zero matrices with implicit dimension, where $e_i:=I(:,i)$. The diagonal matrix with diagonal elements $a_1, \ldots, a_n$ is denoted by $\textrm{diag}(a_1, \ldots, a_n)$. We use standard notation to describe linear systems, such as the quadruple $(A,B,C,D)$ to denote a state-space realization of transfer function $G(s)$, $x(t)$ to describe a time-dependent variable and $X(s)$ its Laplace transform and we omit the dependence on $t$ or $s$ when the meaning is clear. Superscripts are used to highlight particular systems. We also define a \emph{signed identity matrix} as any square, diagonal matrix $J$ that satisfies ${J(i,i) = \pm1}$.

\section{Preliminaries} \label{sec:pre}

\subsection{Spectral Factorization}

Consider systems defined by the following Linear, Time-Invariant (LTI) representation:
\begin{equation}\label{LTI00}
\begin{aligned}
\dot{x} &= Ax + Bu \\
y &= Cx + Du
\end{aligned}
\end{equation}

\noindent with input ${u(t) \in \RR^m}$, state ${x(t) \in \RR^n}$, output ${y(t) \in \RR^p}$, system matrices ${A \in \mathbb{R}^{n \times n}}$, ${B \in \mathbb{R}^{n \times m}}$, ${C \in \mathbb{R}^{p \times n}}$ and ${D \in \mathbb{R}^{p \times m}}$ and transfer function from $u$ to $y$: ${G(s) = C(sI-A)^{-1}B + D}$. Make the following assumptions:
\begin{myass} \label{assA} The matrix $A$ is Hurwitz. \end{myass}
\begin{myass} \label{asse}The system is driven by unknown white noise $u(t)$ with covariance $\mathbb{E}[u(t)u^T(\tau)] = I\delta(t-\tau)$. \end{myass}
\begin{myass} \label{assGM} The system $(A,B,C,D)$ is globally minimal. \end{myass}

\noindent The meaning of Assumption \ref{assGM} is explained below. From $y(t)$, the most information about the system that can be obtained is the output spectral density:
\beqs
\Phi(s) = G(s)G^*(s)
\eeqs

\noindent The spectral factorization problem (see for example \cite{youla61}) is that of obtaining spectral factors $G'(s)$ that satisfy: ${G'G'^* = \Phi}$. Note that the degrees of two minimal solutions may be different; hence make the following definition.

\begin{mydef}[Global Minimality]
For a given spectral density $\Phi(s)$, the globally-minimal degree is the smallest degree of all its spectral factors.
\end{mydef}

\noindent Any system of globally-minimal degree is said to be \emph{globally minimal}. Anderson \cite{anderson69} provides an algebraic characterisation of all realizations of all spectral factors as follows. Given $\Phi(s)$, define the positive-real matrix $Z(s)$ to satisfy:
\beq \label{Zdef}
Z(s) + Z^*(s) = \Phi(s)
\eeq

\noindent Minimal realizations of $Z$ are related to globally-minimal realizations of spectral factors of $\Phi$ by the following lemma.

\begin{mylem}[\cite{anderson69}] \label{andlem}
Let $(A,B_z,C,D_z)$ be a minimal realization of the positive-real matrix $Z(s)$ of \eqref{Zdef}, then the system $(A,B,C,D)$ is a globally-minimal realization of a spectral factor of $\Phi$ if and only if the following equations hold:
\beq \label{eq:and}
\bali
RA^T + AR &= -BB^T\\
RC^T &= B_z - BD^T\\
2D_z &= DD^T
\eali
\eeq

\noindent for some positive-definite and symmetric matrix ${R\in \RR^{n \times n}}$.
\end{mylem}

\noindent This result was used by Glover and Willems \cite{keith74} to provide conditions of equivalence between any two such realizations, which are stated below.

\begin{mylem}[\cite{keith74}] \label{keithlem}
If $(A,B,C,D)$ and $(A',B',C',D')$ are globally-minimal systems, then they have equal output spectral density if and only if:
\begin{subequations} \label{ks}
\begin{align}
A' &= T^{-1}AT \label{k1}\\
C' &= CT \label{k2}\\
SA^T + AS &= -BB^T + TB'B'^TT^T \label{k3}\\
SC^T &= -BD^T + TB'D'^T \label{k4}\\
DD^T &= D'D'^T  \label{k5}
\end{align}
\end{subequations}
\noindent for some invertible $T \in \mathbb{R}^{n \times n}$ and symmetric $S \in \RR^{n \times n}$.
\end{mylem}

\noindent For any two systems that satisfy Lemma \ref{keithlem} for a particular $S$, all additional solutions for this $S$ may be parameterized by Corollary \ref{col_mp}. This is adapted from \cite{keith74} where it was stated for minimum-phase systems.

\begin{mycol} \label{col_mp}
If $(A,B,C,D)$ and $(A',B',C',D')$ satisfy Lemma \ref{keithlem} for a particular $S$, then all systems that also satisfy Lemma \ref{keithlem} with $(A,B,C,D)$ for the same $S$ are given by:
\beq \label{ksmp}
(T'A'T'^{-1}, T'B'U, C'T'^{-1}, D'U)
\eeq

\noindent for some invertible $T' \in \mathbb{R}^{n \times n}$ and orthogonal $U \in \mathbb{R}^{p\times p}$. If $G(s)$ is square and minimum phase (full rank for all $s$ with $\textrm{Re}(s) > 0$), then for $S=0$, \eqref{ksmp} characterises all realizations of minimum-phase spectral factors.
\end{mycol}

\subsection{Structure in LTI Systems} \label{sec_DSF}
We now suppose that there is some unknown underlying system $(A^0, B^0, C^0, D^0)$ with transfer function $G^0$ and we wish to obtain some information about this system from its spectral density $\Phi^0$. Even if $G^0$ is known to be minimum phase, from Corollary \ref{col_mp} it can only be found up to multiplication by some orthogonal matrix $U$. Given $G^0$, the system matrices can also only be found up to some change in state basis. The zero superscript is used to emphasize a particular system.

The following additional assumption is made:
\begin{myass} \label{assDSF} The matrices $C=\bmat I & 0 \emat$ and $D=0$.\end{myass}

\noindent The form of $C$ implies a partitioning of the states into \emph{manifest} variables which are directly observed and \emph{latent} variables which are not. The form of $D$ restricts the systems to be strictly proper, and hence causal. For this class of systems we seek to identify causal dependencies among manifest variables, defined in \cite{TAC08}, as follows.

Partition \eqref{LTI00} under Assumption \ref{assDSF}:
\beq \label{LTI}
\bmat \dot{y}\\ \dot{z} \emat = \bmat A_{11} & A_{12} \\ A_{21} & A_{22} \emat
\bmat y \\ z \emat + \bmat B_1 \\ B_2 \emat u 
\eeq

\noindent where  $y = \bmat I & 0 \emat \bmat y \\ z \emat$ and $z(t) \in \mathbb{R}^l$ are the $l=n-p$ latent states. Taking the Laplace transform of \eqref{LTI} and eliminating $Z$ yields ${sY = W Y + V U}$, for proper transfer matrices:
\beq \label{WVdef}
\bali
W &:= A_{12}\left ( sI - A_{22} \right )^{-1} A_{21} + A_{11}\\
V &:= A_{12}\left (sI - A_{22} \right )^{-1} B_2 + B_1
\eali
\eeq

\noindent Now define $W_D := \textrm{diag}(W(1,1), \ldots, W(p,p))$, subtract $W_DY$ from both sides of ${sY = W Y + V U}$ and rearrange to give:
\begin{equation} \label{YQP}
Y = QY + PU
\end{equation}

\noindent where
\beq \label{QPdef}
\bali
Q &:= \left ( sI - W_D \right )^{-1} \left ( W-W_D \right ) \\
P &:= \left ( sI- W_D \right )^{-1} V
\eali
\eeq

\noindent are strictly-proper transfer matrices of dimension $p\times p$ and $p \times m$ respectively. Note that $Q$ is constructed to have diagonal elements equal to zero (it is hollow).

\begin{mydef}[Dynamical Structure Function]
Given any system \eqref{LTI00} under Assumption \ref{assDSF}, the Dynamical Structure Function (DSF) is defined as the couple $(Q,P)$, where $Q$ and $P$ are given in \eqref{QPdef}.
\end{mydef}

The DSF defines a directed graph with only the manifest states and inputs as nodes. There is an edge from $Y(j)$ to $Y(i)$ if $Q(i,j) \neq 0$; and an edge from $U(j)$ to $Y(i)$ if $P(i,j) \neq 0$. In this sense, the DSF characterises causal relations among manifest states $Y$ and inputs $U$ in system \eqref{LTI00}. The transfer function $G$ is related to the DSF as follows:
\begin{equation} \label{GQP}
G = (I-Q)^{-1}P
\end{equation}

\noindent where, given $G$, the matrices $Q$ and $P$ are not unique in general, hence the following definition is made.
\begin{mydef}[Consistency]
A DSF $(Q,P)$ is consistent with a transfer function $G$ if \eqref{GQP} is satisfied.
\end{mydef}

\begin{figure}[t]
\centering
\includegraphics{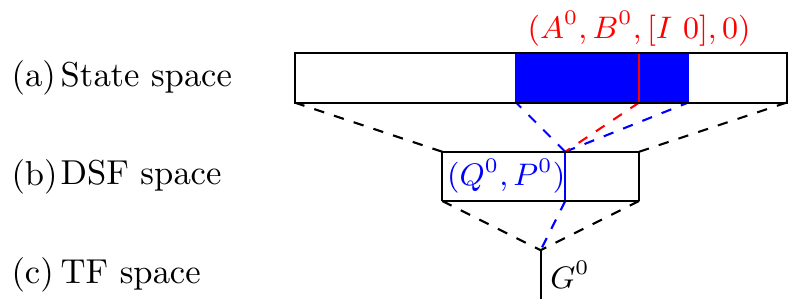}
\caption{Pictorial representation of relationship between state space, DSF space and transfer function space for a particular system $(A^0,B^0,[ I \ 0],0)$. In (a) is contained the set of state transformations of this system by matrices $T$ that preserve $C^0=[I \ 0]$; in red is the particular realization with $T=I$ and in blue the set of realizations with the same DSF $(Q^0,P^0)$. In (b) is the set of all DSFs that have realizations in (a); in blue is the particular DSF $(Q^0,P^0)$. In (c) is the single transfer function $G^0$, with which are consistent all DSFs in (b) and which can be realized by all realizations in (a).}
\label{fig:struc}
\end{figure}

We also define a state-space realization of a particular DSF $(Q^0,P^0)$ as any realization for which the (unique) DSF is $(Q^0,P^0)$. The relationship between state space, DSF and transfer function representations is illustrated in Fig. \ref{fig:struc}, which shows that a state-space realization uniquely defines both a DSF and a transfer function. However, multiple DSFs are consistent with a given transfer function and a given DSF can be realized by multiple state-space realizations.

All realizations of a particular $G$ are parameterized by the set of invertible matrices $T \in \mathbb{R}^{n \times n}$. A subset of these will not change the DSF as follows.
\begin{mydef}[(Q,P)-invariant transformation]
A state transformation $T$ of system $(A^0,B^0,C^0,D^0)$ with DSF $(Q^0,P^0)$ is \emph{$(Q,P)$-invariant} if the transformed system $(TA^0T^{-1},TB^0,C^0T^{-1},D^0)$ also has DSF $(Q^0,P^0)$.
\end{mydef}

\noindent The blue region in Fig. \ref{fig:struc}(a) is the set of all $(Q,P)$-invariant transformations of ${(A^0,B^0,\bmat I & 0 \emat, 0)}$.

\subsection{Network Reconstruction}

The network reconstruction problem was cast in \cite{TAC08} as finding exactly $(Q^0,P^0)$ from $G^0$. Since in general multiple DSFs are consistent with a given transfer function, some additional \emph{a priori} knowledge about the system is required for this problem to be well posed. It is common to assume some knowledge of the structure of $P$, as follows.
\begin{myass} \label{assPdiag}The matrix $P$ is square, diagonal and full rank.\end{myass}

\noindent This is a standard assumption in the literature \cite{TAC08,preci,mati_tac,VdH_autom} and equates to knowing that each of the manifest states is \emph{directly} affected only by one particular input. By direct we mean that there is a link or a path only involving latent states from the input to the manifest state. In the stochastic case considered here, each manifest state is therefore driven by its own \emph{intrinsic} variation. The case in which inputs are also applied to the latent states is considered in Section \ref{sec:full}.

The following theorem is adapted from Corollary 1 of \cite{TAC08}:
\begin{mythe}[\cite{TAC08}] \label{QPcor}
There is at most one DSF $(Q,P)$ with $P$ square, diagonal and full rank that is consistent with a transfer function $G$.
\end{mythe}

\noindent Given a transfer function $G^0$ for which the generating system is known to have $P^0$ square, diagonal and full rank, one can therefore uniquely identify the \lq\lq true\rq\rq \ DSF $(Q^0,P^0)$.

\subsection{Example} \label{eg1}
\begin{figure}[t]
\centering
\includegraphics{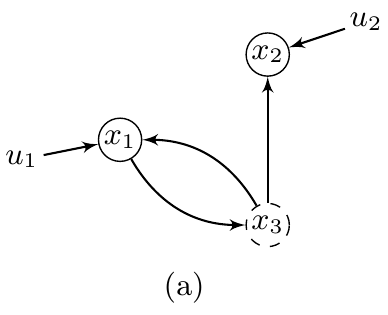}
\includegraphics{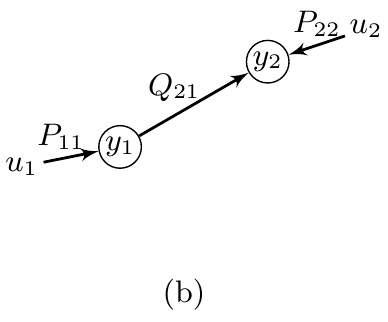}
\caption{Graphical representation of (a) example system $(A^0, B^0, C^0, D^0)$ and (b) its DSF $(Q^0,P^0)$ from Section \ref{eg1}. The DSF describes causal interactions among manifest states that may occur via latent states in the underlying system.}
\label{fig:eg1}
\end{figure}

Consider the following stable, minimal system with two manifest states and one latent state:
\beqs
A^0 = \bmat -1 & 0 & 4 \\ 0 & -2 & 5 \\ -6 & 0 & -3 \emat, \quad B^0 = \bmat 1 & 0 \\ 0 & 1 \\ 0 & 0 \emat,
\eeqs

\noindent with $C^0=\bmat I & 0 \emat$ and $D^0=0$. The system transfer matrix is given by:
\beqs
G^0(s) = \bmat \frac{s+3}{s^2+4s+27} & 0 \\ \frac{-30}{(s+2)(s^2+4s+27)} & \frac{1}{s+2}\emat
\eeqs

\noindent and may be realized by an infinite variety of $A$ and $B$ matrices. The DSF is given by:
\beqs
Q^0(s) = \bmat 0 & 0 \\ \frac{-30}{(s+2)(s+3)} & 0 \emat, \quad P^0(s) = \bmat \frac{s+3}{s^2+4s+27} & 0\\ 0 & \frac{1}{s+2}\emat
\eeqs

\noindent and is the only valid $Q$ and diagonal $P$ that is consistent with $G^0$. This system is represented graphically in Fig. \ref{fig:eg1}.

\subsection{Two realizations for diagonal $P$}

The presence of latent states allows some freedom in the choice of realization used to represent a particular DSF. It will be convenient to use particular forms for systems with $P$ square and diagonal, defined here. We start with the following lemma.

\begin{mylem} \label{Pdiaglem}
The matrix $V$ (and hence $P$) is diagonal if and only if the matrices:
\beqs
B_1 \quad \textrm{and} \quad A_{12}A_{22}^kB_2
\eeqs
\noindent for $k=0,1,\ldots,l-1$ are diagonal, where $l = \textrm{dim}(A_{22})$.
\end{mylem}

\noindent The proof is given in Appendix \ref{Pdiaglem_pf}. Hence, without loss of generality, order the manifest states such that $B_1$ can be partitioned:
\beqs 
B_1 = \bmat 0 & 0 \\ 0 & B_{22} \emat
\eeqs

\noindent where $B_{22}$ is square, diagonal and full rank. Any system $(A^0, B^0, C^0, D^0)$ with $P^0$ square and diagonal can be transformed using $(Q,P)$-invariant transformations into one in the following form. Note that any transformation that preserves $P$ diagonal is $(Q,P)$-invariant by Theorem \ref{QPcor}.

\begin{mydef}[P-Diagonal Form 1] \label{Pcanon}
Any DSF $(Q,P)$ with $P$ square, diagonal and full rank has a realization with $A_{12}$, $A_{22}$, $B_1$ and $B_2$ as follows:
\beq \label{WVcanon}
\left[ \barr {c|c} A_{12} & B_1 \\ \hline A_{22} & B_2 \earr \right] = 
\left[ \barr {cc|cc}
\hc & 0 & 		0 & 0 \\
0 & \times & 			0 & B_{22}\\
\hline
\ha  & \times & 			\hb & 0 \\
0 & \times &			0 & 0 \earr \right]
\eeq

\noindent where $\times$ denotes an unspecified element. The following is a canonical realization of $V = A_{12}\left (sI - A_{22} \right )^{-1} B_2 + B_1$:
\beq \label{Vcanon}
\left(\ha, \bmat \hb & 0 \emat, \bmat \hc \\ 0 \emat, \bmat 0 & 0 \\ 0 & B_{22} \emat \right)
\eeq

\noindent where $\ha := \textrm{diag}(\alpha_{1}, \cdots, \alpha_{p_{11}})$, $\hb := \textrm{diag}(\beta_{1}, \cdots, \beta_{p_{11}})$ and $\hc := \textrm{diag}(\gamma_{1}, \cdots, \gamma_{p_{11}})$, $p_{22} = \textrm{dim}(B_{22})$, $p_{11} = p-p_{22}$ and where $(\alpha_i, \beta_i, \gamma_i, 0)$ is a minimal realization of $V(i,i)$ in controllable canonical form (see \cite{zdg} for example). Denote the dimension of $\al_i$ as $r_i := \din(\al_i)$.\end{mydef}

Further $(Q,P)$-invariant transformations can be applied to systems of the form \eqref{WVcanon} to give a second realization as follows.

\begin{mydef} [P-Diagonal Form 2]
Any DSF $(Q,P)$ with $P$ square, diagonal and full rank has a realization with $A_{12}$, $A_{22}$, $B_1$ and $B_2$ as follows:
\beq \label{Vnmp}
\resizebox{.89\hsize}{!}{$
\left[ \barr {c|c} A_{12} & B_1 \\ \hline A_{22} & B_2 \earr \right] = 
\left[ \barr {cccc|ccc}
0 & 0 & I & 0 & 		0 & 0 & 0\\
0 & \ga_{22} & 0 & 0 & 		0 & 0 & 0\\
\times & \times & \times &  \ga_{34} &		0 & 0 & B_{22}\\
\hline
\times & \times & \times & \al_{14} & 			 I & 0 & 0\\
\times & \times & \times & \al_{24} & 			 0 & I & 0\\
\al_{31} & 0 & \times & \al_{34} &			0 & 0 & 0\\
\times & \times & \times & \al_{44} & 					0 & 0 & 0\earr \right]
$}
\eeq

\noindent where $B_{22} \in \RR^{p_3 \times p_3}$ and $\ga_{22} \in \RR^{p_2 \times p_2}$ are square, diagonal and full rank and $\times$ denotes an unspecified element. The dimension of $B_1$ is ${p = \textrm{dim}(B_1) = p_1+p_2+p_3}$ and the matrix ${\al_{31} \in \RR^{p_1 \times p_1}}$ is square and diagonal but not necessarily full rank. The elements of $A_{22}$ satisfy the following properties for $i = 1, \ldots, p_1$:
\beq
\bali
&\al_{31}(i,:) = \al^{(2)}_{31}(i,:)  = \cdots = \al^{(k_i-1)}_{31}(i,:)  = 0^T\\
&\al^{(k_i)}_{31}(i,:) \neq 0^T\\
&\al^{(k_i)}_{34}(i,:) = 0^T
\eali
\eeq

\noindent for some $1 \leq k_i \leq l-1$, where $l := \textrm{dim}(A_{22})$ and $\al^{(k)}_{ij} := A_{22}^k[i,j]$ denotes block $(i,j)$ of $A_{22}^k$.
\end{mydef}

\noindent The diagonal elements of $V$ have been ordered according to their relative degrees: the first $p_1$ elements have relative degree greater than one; the next $p_2$ have relative degree of one; the last $p_3$ have relative degree of zero. The problem is considerably simpler if all elements have relative degree of zero, in which case $p=p_3$.

\section{Main Results} \label{sec:main}

Given an underlying system $(A^0,B^0,C^0,D^0)$ with DSF $(Q^0,P^0)$, transfer function $G^0$ and output spectral density $\Phi^0$ under Assumptions \ref{assA} - \ref{assPdiag}, we seek to identify $(Q^0,P^0)$ from $\Phi^0$. From Theorem \ref{QPcor}, we know that $(Q^0,P^0)$ can be found uniquely from $G^0$; in Section \ref{ssec:MP} we prove that if $G^0$ is minimum phase, we can find it from $\Phi^0$ up to a choice of sign for each of its columns. From the spectral density $\Phi^0$, we can therefore find $Q^0$ exactly and $P^0$ up to a choice of signs.

Section \ref{ssec:NMP} considers the general case, including non-minimum-phase transfer functions, in which all spectral factors of $\Phi^0$ that satisfy the assumptions can be characterised as solutions to a single Algebraic Riccati Equation (ARE). Necessary and sufficient conditions for two DSFs to have equal spectral factors are given. An algorithm is presented in Section \ref{ssec:alg} to construct all DSF solutions from the spectral density, and illustrated by an example in Section \ref{ssec:eg2}.

\subsection{Minimum-Phase $G$} \label{ssec:MP}

Suppose the following assumption holds:
\begin{myass} \label{assMP} The transfer function $G(s)$ is minimum phase (full rank for all $s$ with $\textrm{Re}(s) > 0$).\end{myass}

\noindent In this case, any two spectral factors $G$ and $G'$ are related by: $G' = GU$ for some orthogonal matrix $U \in \RR^{p \times p}$ \cite{keith74}. We first provide some intuition for $G$ to be minimum phase by the following lemma.
\begin{mylem}
If $Q$ is stable and $P$ is square, diagonal and minimum phase then $G$ is minimum phase.
\end{mylem}

\begin{IEEEproof}
If $Q$ is stable, then $I-Q$ is also stable as it has the same poles. For any invertible transfer function, $z_0$ is a transmission zero if and only if it is a pole of the inverse transfer function \cite{zdg}. Therefore $(I-Q)^{-1}$ is minimum phase if and only if $Q$ is stable. If in addition $P$ is minimum phase, then ${G = (I-Q)^{-1}P}$ is also minimum phase since $P$ is diagonal.
\end{IEEEproof}

\noindent Hence systems with stable interactions among manifest variables that are driven by filtered white noise, where the filters are minimum phase, satisfy Assumption \ref{assMP}.
\begin{mythe} \label{the_mp}
Two systems $(A,B,C,D)$ and $(A',B',C',D')$ under Assumptions \ref{assA}-\ref{assMP} with DSFs $(Q,P)$ and $(Q',P')$
have equal output spectral density:
\beqs
\Phi(s) = G(s)G^*(s) = G'(s)G'^*(s)
\eeqs

\noindent if and only if $G' = GJ$, for some signed identity matrix $J$. This is equivalent to having $Q'=Q$ and $P'=PJ$. Given a particular $\Phi^0$, the minimum-phase spectral factor $G^0J$ is therefore unique up to some choice of $J$, after which the solution for the DSF $(Q^0, P^0J)$ is unique.
\end{mythe}

\begin{IEEEproof}
From Lemma \ref{col_mp}, two systems under Assumptions \ref{assA}-\ref{assPdiag} have equal output spectral density if and only if they satisfy \eqref{ksmp} for some invertible $T\in \mathbb{R}^{n\times n}$ and orthogonal $U\in \mathbb{R}^{p\times p}$. We shall derive necessary conditions for \eqref{ksmp} to hold and show that these imply that $U$ must be a signed identity matrix.

First, $C'=CT^{-1}$ from \eqref{ksmp} is satisfied if and only if $T=\bmat I & 0 \\ T_1 & T_2 \emat$, for some $T_1 \in \RR^{l \times p}$ and invertible $T_2 \in \RR^{l \times l}$. Then $B' = TBU$ gives:
\bseq
\begin{align}
B_1' &= B_1U \label{BU_a}\\
B_2' &= (T_1B_1 + T_2B_2)U \label{BU_b}
\end{align}
\eseq

\noindent Take $(A,B)$ and $(A',B')$ to be in P-diagonal form 1 \eqref{WVcanon}; then from \eqref{BU_a} the size of the partitioning of $B_1$ is the same as that of $B_1'$. Since $B_1'$ must be diagonal, \eqref{BU_a} implies ${U = \bmat U_{11} & 0 \\ 0 & J_{22} \emat}$ partitioned as $B_1$ for some orthogonal $U_{11}$ and signed identity $J_{22}$.

In the case that $B_1$ is invertible ($B_1 = B_{22}$), it is clear that $U=J_{22}$ and the result holds. The result for the general case is the same and the proof given in Appendix \ref{app:pfMP}. We must therefore have $U=J$ for some signed identity matrix $J$ in order for \eqref{ksmp} to be satisfied. From \eqref{GQP}, equality of spectral densities implies:
\beqs
G' = (I-Q')^{-1}P' = (I-Q)^{-1}PJ = GJ
\eeqs

\noindent Inverting the above and equating diagonal elements yields $P' = PJ$ and hence $Q'=Q$.
\end{IEEEproof}

\noindent Given only the spectral density $\Phi^0$, the reconstruction problem for minimum-phase systems therefore has a unique solution for $Q^0$ irrespective of topology. We find this to be a surprising and positive result. The sign ambiguity in $P^0$ is entirely to be expected as only the variance of the noise is known.

\subsection{Non-Minimum-Phase $G$} \label{ssec:NMP}

We now relax Assumption \ref{assMP} to include non-minimum-phase solutions. A straightforward corollary of Theorem \ref{the_mp} is the following.
\begin{mycol} \label{col_SQP}
If two systems $(A,B,C,D)$ and $(A',B',C',D')$ under Assumptions \ref{assA} - \ref{assPdiag} with DSFs $(Q,P)$ and $(Q',P')$ satisfy Lemma \ref{keithlem} for a particular $S$, then all additional systems that also satisfy Lemma \ref{keithlem} with $(A,B,C,D)$ for the same $S$ have DSFs:
\beqs
(Q',P'J)
\eeqs

\noindent for some signed identity matrix $J$. Each solution $S$ to Lemma \ref{keithlem} therefore corresponds to at most one solution for the DSF $(Q',P'J)$ for some choice of $J$.
\end{mycol}

\noindent Next we prove that for a given system $(A,B,C,D)$, solutions $S$ to Lemma \ref{keithlem} can be partitioned into two parts: the first must be zero and the second must solve an Algebraic Riccati Equation (ARE) with parameters determined by the original system.

Analagous to the minimum-phase case, we evaluate solutions to Lemma \ref{keithlem} for any two realizations that satisfy Assumptions \ref{assA} - \ref{assPdiag} and are in P-diagonal form 2 \eqref{Vnmp}. Since every such system can be realized in this form, these results are completely general given the assumptions made. Immediately \eqref{ks} yields:
\bseq \label{eqgen}
\begin{align}
A' ={}& T^{-1}AT \label{gl1}\\
B_{1}'B_{1}'^T ={}& B_{1}B_{1}^T \label{gl2}\\
A_{12}S_{2} ={}& B_1B_1^TT_1^T \label{gl3}\\
S_2A_{22}^T + A_{22}S_2 ={}& -B_2B_2^T + T_1B_1B_1^TT_1^T \label{gl4}\\
& + T_2B_2B_2^TT_2^T\nonumber
\end{align}
\eseq

\noindent where
\beq\label{ST}
S = \bmat 0 & 0 \\ 0 & S_2 \emat \qquad \mathrm{and} \qquad T = \bmat I & 0 \\ T_1 & T_2 \emat
\eeq

\noindent with $S_2 \in \RR^{l \times l}$, $T_1 \in \RR^{l \times p}$ and $T_2 \in \RR^{l \times l}$. We can further partition $S_2$ as follows.

\begin{mylem}\label{S2lem}
For any two systems $(A,B,C,D)$ and $(A',B',C',D')$ in P-diagonal form 2 that satisfy Assumptions \ref{assA}-\ref{assPdiag} and Lemma \ref{keithlem}, the matrix $S_2$ in \eqref{ST} satisfies:
\beqs \label{S22}
S_2 = \bmat  0 & 0 \\ 0 & s_{22} \emat
\eeqs

\noindent where $s_{22} \in \RR^{l_2 \times l_2}$ and $l_2 := l-2p_1 - p_2$.
\end{mylem}

The proof is given in Appendix \ref{app:pf_s2} and is obvious if $B_1$ is invertible, in which case $l=l_2$. The above lemma significantly simplifies \eqref{eqgen}. Whilst there is some freedom in the choice of $T_1$ and $T_2$, the number of solutions for $s_{22}$ only depends on the system in question, as follows.

\begin{mythe} \label{mainth}
Two DSFs $(Q,P)$ and $(Q',P')$ with realizations $(A,B,C,D)$ and $(A',B',C',D')$ in P-diagonal form 2 under Assumptions \ref{assA}-\ref{assPdiag} have equal output spectral density if and only if the following equations are satisfied:
\beq \label{are}
s_{22}\bar{a} + \bar{a}^Ts_{22} - s_{22}\bar{b}\bar{b}^T s_{22} = 0
\eeq
\beq \label{a34s22}
\al_{34}s_{22} = 0 
\eeq

\noindent for some symmetric $s_{22} \in \RR^{l_2 \times l_2}$ with ${l_2=l-2p_1 - p_2}$, where $\bar{a}$ and $\bar{b}$ are comprised of parameters of $A$ and $B$ in \eqref{Vnmp} as $\bar{a} :=\al_{44}^T$ and ${\bar{b} := \bmat \ga_{34}^TB_{22}^{-T} & \al_{14}^T & \al_{24}^T & \al_{34}^T \emat}$ and there exists invertible ${T = \bmat I & 0 \\ T_1 & T_2 \emat}$ such that:
\bseq \label{pf_rest}
\begin{align}
A' &= T^{-1}AT\\
B_{22}' &= B_{22}J\\
T_1B_1 &= \bmat 0 & 0 \\ 0 & 0 \\ 0 & s_{22}\ga_{34}^TB_{22}\emat \\
T_2B_2 &= \bmat t_{1} & 0 \\ 0 & 0 \\ s_{22} \bmat \al_{14}^T & \al_{24}^T \emat t_1 & 0\emat
\end{align}
\eseq

\noindent for some orthogonal ${t_1 \in \RR^{(p_1+p_2) \times (p_1 + p_2)}}$ and signed identity matrix $J \in \RR^{p_3\times p_3}$.
\end{mythe}

\begin{IEEEproof}
The proof follows directly from Lemma \ref{S2lem} by substituting $S_2 = \bmat 0 & 0 \\ 0 & s_{22} \emat$ into \eqref{eqgen}.
\end{IEEEproof}

\begin{myrem}
From Corollary \ref{col_SQP}, the number of DSFs that have equal spectral density to that of any given $(Q^0,P^0)$ (up to a choice of signed identity matrix) is therefore at most equal to the number of solutions to the ARE \eqref{are}.
\end{myrem}

\begin{myrem}
It is straightforward to see that $(\bar{a}, \bar{b})$ is controllable due to the minimality of $(A, B, C, D)$. The number of solutions to \eqref{are} can therefore be calculated from the Hamiltonian matrix of \eqref{are} and in particular is finite if and only if every eigenvalue has unit geometric multiplicity (see \cite{lancaster}). In general the solution will not be unique.
\end{myrem}

\begin{myrem}
Any solution to the ARE must also satisfy \eqref{a34s22} in order to satisfy Theorem \ref{mainth}. This condition will not necessarily be satisfied, reducing the size of the DSF solution set.
\end{myrem}

\begin{myrem} \label{rem:TJ}
Given any system $(A^0,B^0,C^0,D^0)$ with DSF $(Q^0,P^0)$, the solution set of \eqref{are} can be calculated; for any solution that also satisfies \eqref{a34s22} it is straightforward to choose $T$ and $J$ to satisfy \eqref{pf_rest}; however, the resulting transformed system may not have $P$ diagonal, and it is nontrivial to choose $T$ such that it does.
\end{myrem}

\begin{myrem} \label{rem:l2}
The $l$ latent states have been partitioned into $l_2 = l-2p_1-p_2$ and $l_1 = l-l_2$, where the dimension of the ARE \eqref{are} is $l_2$. The number of DSF solutions is therefore principally determined by $l_2$ and not by $p$, the number of measured states; a ``large'' network (high $p$) could have relatively few spectrally-equivalent solutions if it has ``small'' $l_2$. Any system with $l_2=0$ has a unique solution.
\end{myrem}

\subsection{A Constructive Algorithm} \label{ssec:alg}

If $B_1$ is invertible, the transfer functions $V(i,i)$ have relative degree of zero and any realization in P-diagonal form 2 has $B_2=0$. Any system with $B_1$ diagonal and $B_2=0$ trivially satisfies Lemma \ref{Pdiaglem} and hence has $P$ diagonal. In this case, the matrix $T_1$ is completely determined by \eqref{pf_rest} and $T_2$ can be chosen freely. For this class of systems, DSFs $(Q,P)$ with output spectral density $\Phi(s)$ can be constructed. A procedure is outlined below, which essentially finds solutions to Lemma \ref{andlem} under Assumptions \ref{assA}-\ref{assPdiag}.

\begin{enumerate}
\item Estimate $\Phi(s)$
\item Construct positive-real $Z(s)$ such that: ${Z+Z^*=\Phi}$
\item Make a minimal realization of $Z$ of the form: ${(A,B,\bmat I & 0 \emat,0)}$
\item Find all solutions to the following equations for $B'$ and symmetric $R>0$:
\begin{subequations} \label{QPsolve}
\begin{align}
RA^T + AR&=-B'B'^T \label{s1}\\
RC^T&=B \label{s2}
\end{align}
\end{subequations}
\item For each solution, $B'$, the system ${(A,B',\bmat I & 0 \emat,0)}$ with transfer function $G'$ is minimal and has spectral density $\Phi$ from Lemma \ref{andlem}
\item Apply the state transformation $T = \bmat I & 0 \\ -B_2'B_1'^{-1} & I \emat$ to obtain system  ${(A'',B'',\bmat I & 0 \emat,0)}$, which has the same transfer function $G'$, spectral density $\Phi$ and has $P''$ diagonal. The system $(Q'',P'')$ then satisfies Assumptions \ref{assA}-\ref{assPdiag} and has spectral density $\Phi$.
\end{enumerate}

\noindent Step 1 may be achieved by standard methods, the details of which are not considered here. Every spectral density matrix has a decomposition of the form of Step 2, as described in \cite{anderson69} and Step 3 is always possible for strictly-proper $Z$. Solutions to step 4 can be obtained as follows.

Equation \eqref{QPsolve} derives from \eqref{eq:and} under Assumption \ref{assDSF}. Partitioning \eqref{s2} gives:
\beq
R = \bmat B_1 & B_2^T\\ B_2 & R_2 \emat
\eeq

\noindent for some symmetric $R_2$. Equation \eqref{s1} then defines three equations:
\begin{subequations} \label{eqsolv}
\begin{align}
&B_1'B_1'^T = -B_1A_{11}^T - B_2^TA_{12}^T - A_{11}B_1 - A_{12}B_2 \label{es1}\\
&R_2A_{12}^T = -B_2A_{11}^T - A_{21}B_1 - A_{22}B_2 - B_2'B_1'^T \label{es2}\\
&R_2A_{22}^T + A_{22}R_2 + B_2A_{21}^T + A_{21}B_2^T + B_2'B_2'^T =0 \label{es3}
\end{align}
\end{subequations}

\noindent Both sides of \eqref{es1} must be diagonal and full rank, since such a solution for $B_1'$ is known to exist; hence a diagonal, full-rank solution for $B_1'$ can be found from \eqref{es1}. Given $B_1'$, $B_2'$ can be eliminated from \eqref{es3} using \eqref{es2}, yielding the following ARE in $R_2$:
\beq \label{sARE}
R_2\bar{A} + \bar{A}^TR_2 + R_2\bar{B}\bar{B}^TR_2 + \bar{Q} = 0
\eeq

\noindent with:
\beq
\bali
\bar{A} &= \left(A_{22} + FB_1'^{-2}A_{12}\right)^T\\
\bar{B} &= \left(B_1'^{-1}A_{12}\right)^T\\
\bar{Q} &= B_2A_{21}^T + A_{21}B_2^T + FF^T
\eali
\eeq

\noindent and $F = B_2A_{11}^T + A_{21}B_1 + A_{22}B_2$. Since the parameters of \eqref{sARE} are known, we can compute all symmetric, positive-definite solutions $R_2$. Given $R_2$, the matrix $B_2'$ is given uniquely by \eqref{es2} as:
\beq \label{sB2}
B_2' = -\left(R_2A_{12}^T + B_2A_{11}^T + A_{21}B_1 + A_{22}B_2\right)B_1'^{-1}
\eeq

The system $(A,B',\bmat I & 0 \emat,0)$ with DSF denoted $(Q',P')$ therefore has spectral density $\Phi$ but will in general not have $P'$ diagonal. The transformation of Step 6 results in $B''_2=0$ and hence the transformed $P''$ diagonal from Lemma \ref{Pdiaglem}. Since state transformations do not affect the spectral density, the system $(Q'',P'')$ satisfies Assumptions \ref{assA}-\ref{assPdiag} and has spectral density $\Phi$.

\section{Examples} \label{sec:sim}

\subsection{Example with Two Solutions} \label{ssec:eg2}

Given the output spectral density $\Phi(s)$ for the system of Section \ref{eg1}, we construct all DSF solutions with this spectral density as described in Section \ref{ssec:alg}. From the output spectral density construct the positive-real matrix $Z(s)$, such that $Z(s)+Z^*(s)=\Phi(s)$:
\beqs
Z(s) = \bmat \frac{0.17(s+1)}{s^2+4s+27} & \frac{0.032(s+19)}{s^2+4s+27} \\
\frac{0.032(s+8.2)(s-17.2)}{(s+2)(s^2+4s+27)} & \frac{0.57s^2+2.9s+15.1}{(s+2)(s^2+4s+27)}\emat
\eeqs

\noindent after numerical rounding. Construct any minimal realization of $Z$ by standard methods, such as:
\beqs
A = \bmat -3.9 & -0.97 & 1.9 \\ -3.6 & -3.2 & 2.4 \\ -15.5 & -1.5 & 1.1 \emat,
\qquad B = \bmat 0.17 & 0.032 \\ 0.032 & 0.57 \\ 0.092 & 0.60 \emat,
\eeqs

\noindent with $C=\bmat I & 0 \emat$ and $D=0$. First solve for $B_1'$ as in \eqref{es1}:
\beq \label{egB1}
B_1' = \bmat \pm 1 & 0 \\ 0 & \pm 1 \emat
\eeq

\noindent and choose the signs to be positive for simplicity. Next construct and solve the ARE \eqref{sARE}, which has the following two solutions:
\beqs
R_2 = 1.02 \quad \textrm{and} \quad 1.65
\eeqs

\noindent In each case, solve for $B_2'$ using \eqref{sB2}:
\beqs
B_2' = \bmat 1.49 & 0.5 \emat \quad \textrm{and} \quad \bmat 0.28 & -1.01 \emat
\eeqs

\noindent and transform both systems by $T = \bmat I & 0 \\ -B_2'B_1'^{-1} & I \emat$ to yield two systems with DSFs with $P$ diagonal. The first corresponds to the system of Example \ref{eg1} with DSF:
\beqs
Q(s) = \bmat 0 & 0 \\ \frac{-30}{(s+2)(s+3)} & 0 \emat, \quad P(s) = \bmat \frac{s+3}{s^2+4s+27} & 0\\ 0 & \frac{1}{s+2}\emat
\eeqs

\noindent and the second to the following stable, minimal system:
\beqs
A' = \bmat -3.3 & -2.9 & 4 \\ -2.9 & -5.7 & 5 \\ -8.3 & -3.7 & 3 \emat, \quad B' = \bmat 1 & 0 \\ 0 & 1 \\ 0 & 0 \emat,
\eeqs

\noindent with $C'=\bmat I & 0 \emat$, $D'=0$ and DSF:
\beqs
Q'(s) = \bmat 0 & \frac{-2.9(s+2.0)}{s^2 + 0.34s + 23.3} \\ \frac{-2.9(s+11.1)}{s^2 + 2.7s + 1.3} & 0 \emat,
\eeqs
\beqs
P'(s) = \bmat \frac{s-3}{s^2 + 0.34s + 23.3} & 0 \\ 0 & \frac{s-3}{s^2 + 2.659s +1.3} \emat
\eeqs

\begin{figure}[t]
\centering
\includegraphics{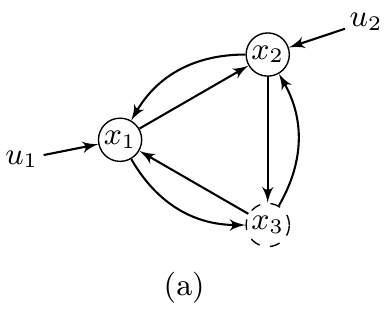}
\includegraphics{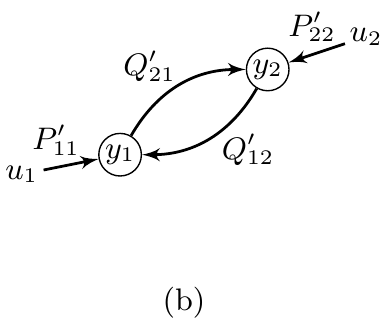}
\caption{Graphical representation of (a) example system $(A', B', C', D')$ and (b) its DSF $(Q',P')$ from Section \ref{ssec:eg2}.}
\label{fig:eg2}
\end{figure}

\noindent Note that this system has a different network structure for both state-space and DSF, as illustrated in Fig. \ref{fig:eg2}.
The reader may verify that these systems do indeed have the same output spectral density $\Phi(s)$. It may also be verified that \eqref{eqgen}, or equivalently Lemma \ref{keithlem}, is  satisfied for the following matrices $S$ and $T$:
\beqs
S = \bmat 0 & 0 & 0 \\ 0 & 0 & 0 \\ 0 & 0 & -0.15 \emat, \quad T = \bmat 1 & 0 & 0 \\ 0 & 1 & 0 \\ -0.59 & -0.73 & 1 \emat
\eeqs

\noindent The transfer matrix for the second system is given by:
\beqs
G'(s) = \bmat \frac{s+0.66}{s^2+4s+27} & \frac{-2.9}{s^2+4s+27}\\ \frac{-2.9(s+11.1)}{(s+2)(s^2+4s+27)} & \frac{s^2+0.34s+23.3}{(s+2)(s^2+4s+27)} \emat
\eeqs

%

\noindent which (from Theorem \ref{the_mp}) is necessarily non-minimum phase -- it has a transmission zero at $s=3$. In this example these are the only two globally-minimal solutions for $Q$ that have diagonal $P$; additional solutions for $P$ may be obtained by changing the signs of $B_1'$ in \eqref{egB1}.

\subsection{Numerical Simulations}

In Remark \ref{rem:l2} it was noted that the principal system dimension that determines the number of DSF solutions is $l_2 = l-2p_1-p_2$ as this is the dimension of the ARE. We simulated 18129 random systems with dimensions in the ranges given in Table \ref{tab:sim}. For each case, random matrices $A$ and $B$ are generated such that the system $(A,B,\bmat I & 0 \emat, 0)$ is stable, minimal and has DSF with $P$ square, diagonal and full rank. Invertible state transformations are then applied to this system to convert it into P-diagonal form 2 and the ARE \eqref{are} is formed.

The number of real, symmetric solutions to \eqref{are} is determined and, if finite, all solutions are constructed. For each solution, \eqref{a34s22} is checked and if satisfied, matrices $T$ and $J$ are chosen to satisfy \eqref{pf_rest}, resulting in a transformed system $(A',B',\bmat I & 0 \emat, 0)$ with DSF $(Q',P')$ and equal spectral density to the original system. As mentioned in Remark \ref{rem:TJ}, the matrix $P'$ may or may not be diagonal for the chosen $T$; if it is not, there may still exist a $T$ for which it is.

\begin{figure}[t]
\centering
\includegraphics[scale = 0.8]{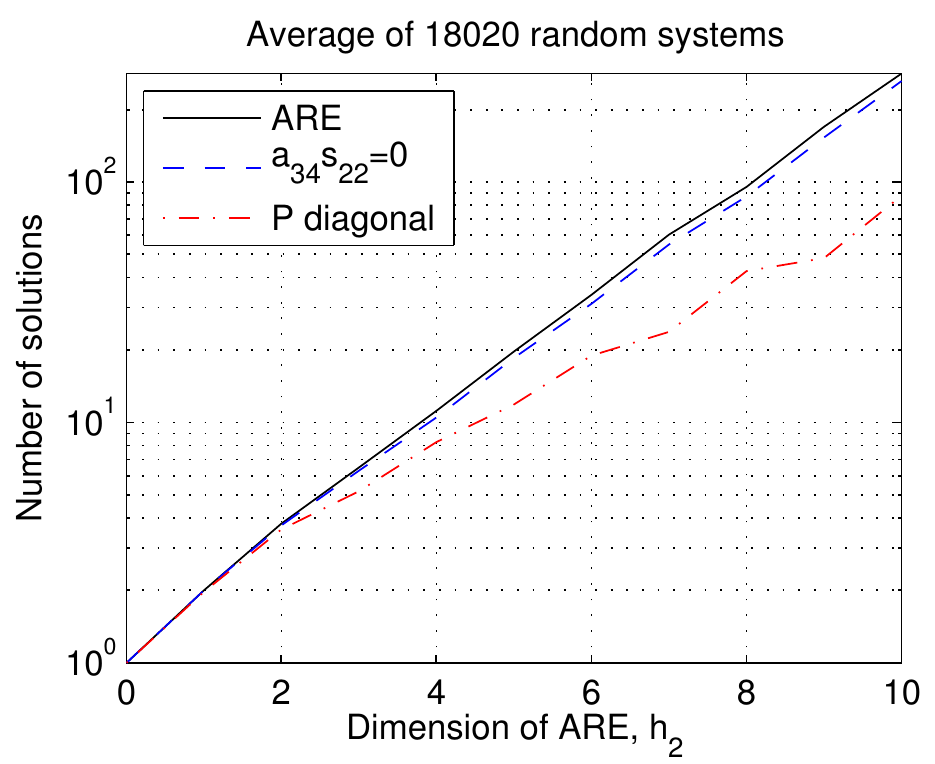}
\caption{Average number of solutions for different cases. In black (-) is the number of solutions to the ARE \eqref{are}; in blue (- -) is the number of solutions that also satisfied \eqref{a34s22}; in red (-.-) is the number of solutions for which a DSF with $P$ diagonal was found.}
\label{fig:sim}
\end{figure}

Of the systems considered, 109 (approximately 0.6\%) resulted in an ARE with a continuum of solutions and hence an infinite number of solutions to the network reconstruction problem. The average number of solutions for the remaining 18020 are shown in Fig. \ref{fig:sim}, from which it can be seen that little restriction is provided by \eqref{a34s22}. The number of solutions found with $P$ diagonal is a lower bound on the actual number of such solutions, which therefore lies between the red and blue lines.

\section{Full Intrinsic Noise} \label{sec:full}

We now relax Assumption \ref{assPdiag} and assume instead the following form of $B$:
\begin{myass}\label{assfull}
The matrix $B$ is given by: $B = \bmat B_{11} & 0 \\ 0 & B_{22} \emat$, where $B_{11} \in \RR^{p \times p}$ and $B_{22} \in \RR^{l \times l}$ are square, full rank and diagonal.
\end{myass}

\noindent This corresponds to the more general scenario of each state being driven by an indepedent noise source. The following lemma, derived directly from Lemma \ref{keithlem}, characterises all DSFs with equal output spectral density.

\begin{mylem} \label{intlem}
Two systems $(A,B,C,D)$ and $(A',B',C',D')$ with DSFs $(Q,P)$ and $(Q',P')$ under Assumptions \ref{assA}-\ref{assDSF} and Assumption \ref{assfull} have equal output spectral density if and only if there exists invertible ${T = \bmat I & 0 \\ T_1 & T_2 \emat \in \mathbb{R}^{n \times n}}$ and symmetric $S_2 \in \RR^{l \times l}$ such that:
\bseq \label{eqint}
\begin{align}
&A' = T^{-1}AT \label{igl1}\\
&B_{11}'B_{11}'^T = B_{11}B_{11}^T \label{igl2}\\
&T_1 = S_{2}A_{12}^TB_{11}^{-2} \label{igl3}\\
&S_2\bar{A} + \bar{A}^TS_2 - S_2\bar{B}\bar{B}^TS_2 + \bar{Q} = 0 \label{igl4}
\end{align}
\eseq

\noindent where $\bar{A}$, $\bar{B}$ and $\bar{Q}$ are defined as:
\beq \label{AREbarint}
\bali
\bar{A} &:= A_{22}^T \\
\bar{B} &:= (B_{11}^{-1}A_{12})^T\\
\bar{Q} &:= B_{22}B_{22}^T - T_2B_{22}'B_{22}'^TT_2^T
\eali
\eeq
\end{mylem}

\noindent It is straightforward to construct multiple solutions to Lemma \ref{intlem} as illustrated by the following example.

\begin{table}[t] 
\renewcommand{\arraystretch}{1.3}
\centering
\caption{Range of system dimension used in simulations}
\begin{tabular}{| c | c | c |}
\hline
Dimension & min & max \\ \hline
$l_2$ & 0 & 10 \\ \hline
$p$ & 2 & 6 \\ \hline
$l$ & 0 & 16 \\ \hline
\end{tabular}
\label{tab:sim}
\end{table}

\subsection{Example with Continuum of Solutions} \label{ssec:egfull}

Consider the system from Example \ref{eg1} with noise inputs applied to every state:
\beqs
A = \bmat -1 & 0 & 4 \\ 0 & -2 & 5 \\ -6 & 0 & -3 \emat, \quad B = \bmat 1 & 0 & 0\\ 0 & 1 & 0 \\ 0 & 0 & 1 \emat
\eeqs

\noindent with $C=\bmat I & 0 \emat$ and $D=0$. The DSF is now given by:
\beqs
Q(s) = \bmat 0 & 0 \\ \frac{-30}{(s+2)(s+3)} & 0 \emat,
\eeqs
\beqs
P(s) = \bmat \frac{s+3}{s^2+4s+27} & 0 & \frac{4}{s^2+4s+27}\\ 0 & \frac{1}{s+2} & \frac{5}{(s+2)(s+3)}\emat
\eeqs

\noindent From \eqref{eqint} the matrices $S$ and $T$ of Lemma \ref{keithlem} are parametrized by a scalar $\theta \in \RR$:
\beqs
S(\theta) = \bmat 0 & 0 & 0 \\ 0 & 0 & 0 \\ 0 & 0 & \theta \emat, \quad T(\theta) = \bmat 1 & 0 & 0 \\ 0 & 1 & 0 \\ 4\theta & 5\theta & T_2(\theta) \emat
\eeqs

\noindent where $S_2=\theta$ and $T_2(\theta) = \sqrt{1-6\theta - 41\theta^2}$. The set ${\{ \theta \ | \ \theta > -0.99 \}}$ defines a continuum of solutions $S(\theta)$ and $T(\theta)$ to Lemma \ref{intlem}, with, for example, $B_{11}' = B_{11}$. This results in a continuum of DSF solutions with equal spectral density in which $Q(\theta)=$
\beqs
\bmat 0 & \frac{20\theta s + 40\theta}{s^2 + (4+25\theta)s + (27 + 25\theta - 592\theta^2)} \\
\frac{20\theta s + (30 - 20\theta - 740 \theta^2)}{s^2 + (5+16\theta)s + (6+32\theta)} & 0 \emat
\eeqs

\noindent Clearly neither the structure nor dynamics of the network are unique.

In practice, therefore, if the commonly-made assumption that noise is only applied at the measured states does not hold, the network reconstruction problem is unlikely to have a unique solution. It may also be verified that this result remains the same even if the network contains no feedback, for example by removing element $(1,3)$ of $A$.

\section{Discussion and Conclusions} \label{sec:conc}
The identifiability of the structure and dynamics of an unknown network driven by unknown noise has been assessed based on factorizations of the output spectral density. Two noise models are considered: noise applied only to the manifest states and noise applied to all the states, the latter of which is shown to be not identifiable in general. For the former noise model, the minimum-phase spectral factor is shown to be unique up to sign and hence such networks are identifiable. Non-minimum-phase spectral factors then correspond to solutions to an Algebraic Riccati Equation, which can be solved to compute all non-minimum-phase networks. The results apply with no restrictions on the topology of the network and can be derived analogously in discrete time.

The development of an efficient estimation algorithm remains a significant challenge. One approach is suggested in which factorizations are made from an estimated spectral density matrix; however robustness to uncertainty in the spectral density is likely to present difficulties. Methods to estimate the network solution directly are currently being considered, one issue being enforcing the requirement for the system to have a minimal realization. Non-minimal solutions are likely to be non-unique.

\appendices

\section{Proof of Lemma \ref{Pdiaglem}} \label{Pdiaglem_pf}
\begin{IEEEproof}
The matrix $P$ is defined in \eqref{QPdef} and is hence diagonal if and only if $V$ is; from the definition of $V$ in \eqref{WVdef}, let $s\rightarrow \infty$ to prove the $B_1$ part. With $B_1$ diagonal, $V$ is diagonal if and only if:
\beq \label{Vdyn}
A_{12}(sI-A_{22})^{-1}B_2
\eeq

\noindent is diagonal. The inverse $(sI-A_{22})^{-1}$ can be expressed as a Neumann series:
\beqs
(sI-A_{22})^{-1} = \frac{1}{s}\left(I-\frac{A_{22}}{s}\right)^{-1} = \frac{1}{s}\sum_{k=0}^{\infty}\left(\frac{A_{22}}{s}\right)^k
\eeqs

\noindent if the sum converges. For $s > \max(|\lambda_i|)$, where $\lambda_i$ are the eigenvalues of $A_{22}$, this requirement is always met and hence \eqref{Vdyn} can be written:
\beqs
A_{12}(sI-A_{22})^{-1}B_2 = \sum_{k=0}^{\infty}\frac{A_{12}A_{22}^kB_2}{s^{k+1}}
\eeqs

\noindent which is a polynomial in $\frac{1}{s}$. This is diagonal if and only if all of its coefficients are, and by the Cayley-Hamilton Theorem it is sufficient to check only the first $l=\textrm{dim}(A_{22})$ of these.
\end{IEEEproof}

\section{Proof of Theorem \ref{the_mp} for $B_1$ not invertible} \label{app:pfMP}

Continuing from the proof in the text, we have that ${U = \bmat U_{11} & 0 \\ 0 & J_{22} \emat}$ for some orthogonal $U_{11}$ and signed identity $J_{22}$; we now show that $U_{11}$ must also be a signed identity matrix. Since the second block column of $B_2'$ must be zero from \eqref{WVcanon}, we require: $T_1B_1=0$. From \eqref{ksmp} we now have:
\bseq
\begin{align}
A_{12}' &= A_{12}T_2^{-1}\\
A_{22}' &= T_2(A_{22} + T_2^{-1}T_1A_{12})T_2^{-1}\\
B_2' &= T_2B_2U
\end{align}
\eseq

\noindent Define $\hat{T_1} := T_2^{-1}T_1$ for clarity. The matrices $V$ and $V'$ are both required to be diagonal by Assumption \ref{assPdiag}, where:
\beqs
\bali
V' &= A_{12}'(sI-A_{22}')^{-1}B_2' + B_1'\\
&= A_{12}(sI-A_{22} - \hat{T_1}A_{12})^{-1}B_2U + B_1J_{22}
\eali
\eeqs

\noindent From Lemma \ref{Pdiaglem}, since $B_1J_{22}$ is diagonal, $V'$ is diagonal if and only if
\beq \label{V_diag}
A_{12}(A_{22} + \hat{T}_1A_{12})^kB_2U
\eeq

\noindent is diagonal for ${k=0,\ldots,l-1}$ where $l=\din(A_{22})$. Given that $V$ is diagonal, we will now show that a necessary condition for $V'$ to be diagonal is that $U_{11}$ is a signed identity. First note that in P-diagonal form 1:
\beq \label{k_term}
A_{12}(A_{22})^kB_2U = \bmat \hc\ha^k\hb U_{11} & 0 \\ 0 & 0 \emat
\eeq

\noindent and
\beq \label{cab}
\hc\ha^k\hb = \diag(\gamma_{1}\alpha_{1}^k\beta_{1}, \ \ldots \ , \gamma_{p_{11}}\alpha_{p_{11}}^k\beta_{p_{11}})
\eeq

\noindent where $p_{11}=\din(U_{11})$. Note also that if ${\gamma_{i,1} = \ldots = \gamma_{i,k} = 0}$:
\beq \label{cabi}
\gamma_{i}\alpha_{i}^k\beta_{i} = \gamma_{i,k+1}
\eeq

\noindent since $(\alpha_i, \beta_i, \gamma_i, 0)$ is a realization in controllable canonical form. Recall that $r_i = \din(\alpha_i)$ is the order of the transfer function $V(i,i)$.

We now prove by induction that the following statement holds for $i = 1, \ldots, p_{11}$ and for all ${k\geq 0}$ if \eqref{V_diag} is diagonal:
\beq \label{kstat}
\boxed{
\begin{array}{ll}
\textbf{If} & \left\{ k < r_i \textbf{ \ and \ } \gamma_{i,1} = \ldots = \gamma_{i,k+1} = 0 \right\}\\
& \left(A_{12}A_{22}^kB_2U\right)(:,i) = 0 \\
& \left(A_{12}A_{22}^kB_2U\right)(i,:) = 0^T\\
\textbf{Else} & \\
& \ U_{11}(:,i) = \left(U_{11}(i,:)\right) ^T = \pm e_i\\
\end{array}
}
\eeq

\subsection*{Base case: ${k=0}$}

\noindent Note that ${r_i > 0}$ for  ${i=1,\ldots,p_{11}}$ (otherwise ${V(i,i)=0}$) and hence $k<r_i$ for $k=0$. Then \eqref{V_diag} requires $A_{12}B_2U$ to be diagonal, which is equivalent to:
\beq \label{k0}
\hc\hb U_{11} = \diag(\gamma_{1,1}, \ldots, \gamma_{p_{11},1})U_{11}
\eeq

\noindent being diagonal. Consider row $i$: if $\gamma_{i,1} = 0$ then clearly $\left(A_{12}B_2U\right)(i,:) = 0^T$ and (since it is diagonal) ${\left(A_{12}B_2U\right)(:,i) = 0}$ too. Conversely, if $\gamma_{i1} \neq 0$ then ${U_{11}(i,j) = 0 \ \ \forall j \neq i}$ (since \eqref{k0} must be diagonal) and therefore ${U_{11}(i,i) = \pm 1}$ such that $U$ is orthogonal. We therefore have ${U_{11}(i,:) = \pm e_i^T}$ and (by a standard property of orthogonal matrices) $U_{11}(:,i) = U_{11}(i,:)^T$. Hence \eqref{kstat} holds for $k=0$ for all $i = 1, \ldots, p_{11}$.

\subsection*{Induction}

\noindent Assume \eqref{kstat} holds for $k = 0, \ldots, \hk-1$ for some $\hk$, for $i = 1, \ldots, p_{11}$, and show that if the $\hk^{th}$ term of \eqref{V_diag} is diagonal then \eqref{kstat} holds for $k=\hk$. The $\hk^{th}$ term of \eqref{V_diag} is: ${A_{12}(A_{22} + \hat{T}_1A_{12})^{\hk} B_2U =}$
\beq \label{kstatsteps}
A_{12}A_{22}^{\hk} B_2U  \quad + \quad \sum_{h=0}^{\hk-1}\zeta_h A_{12}A_{22}^hB_2U
\eeq

\noindent for some $\zeta_h$. Consider any $i$ for which ${\gamma_{i,1} = \ldots = \gamma_{i,\hk} = 0}$ and note that $\gamma_{i,\hk+1} =0 \ \Rightarrow \ \hk<r_i$, otherwise $V(i,i)=0$. Hence we must show (a) that if $\gamma_{i,\hk+1} = 0$, the $i^{th}$ column and row of $A_{12}A_{22}^{\hk} B_2U$ are zero vectors and (b) that if $\gamma_{i,\hk+1}\neq0$, the $i^{th}$ column and row of $U_{11}$ are (signed) unit vectors. 

\subsubsection*{(a)}
Suppose $\gamma_{i,\hk+1} = 0$, then from \eqref{k_term} and \eqref{cabi}:
\beqs
\left(A_{12}A_{22}^{\hk} B_2U\right)(i,:) = \bmat \gamma_{i,\hk+1}U_{11}(i,:) & 0^T \emat = 0^T
\eeqs

\noindent as desired. To show $\left(A_{12}A_{22}^{\hk} B_2U\right)(:,i) = 0$ note that the $i^{th}$ column of the second term in \eqref{kstatsteps} is zero from \eqref{kstat} for $k = 0, \ldots, \hk-1$ and hence the $i^{th}$ column of $A_{12}A_{22}^{\hk} B_2U$ must also be zero such that \eqref{kstatsteps} is diagonal.

\subsubsection*{(b)}
Suppose $\gamma_{i,\hk+1} \neq 0$ and consider element $(i,j)$ of \eqref{kstatsteps} for $j\neq i$, which must be equal to zero. If ${\gamma_{j,1} = \ldots = \gamma_{j,\hk} = 0}$ then from \eqref{kstat} for $k = 0, \ldots, \hk-1$, the $j^{th}$ column of the second term in \eqref{kstatsteps} is zero and element $(i,j)$ is determined only by $A_{12}A_{22}^{\hk} B_2U$, giving:
\beqs
\left(\hc\ha^{\hk}\hb U_{11}\right)(i,j) = \gamma_{i,\hk+1}U_{11}(i,j)=0
\eeqs

\noindent and hence $U_{11}(i,j) = 0$. Otherwise, if $\gamma_{j,h} \neq 0$ for some $1\leq h \leq \hk$, we have $U_{11}(:,j) = \pm e_j$ and hence $U_{11}(i,j) = 0$ directly from \eqref{kstat}. Therefore ${U_{11}(i,j) = 0 \ \ \forall j \neq i}$ which requires ${U_{11}(i,:) = \pm e_i^T = U_{11}(:,i)^T}$ as desired.

\subsection*{Termination}

\noindent Therefore by induction \eqref{kstat} holds for $i = 1, \ldots, p_{11}$ for all $k\geq0$. In particular, it holds for $k=\max_i(r_i)$, in which case the \lq\lq if\rq\rq \ condition is never satisfied and ${U_{11}(:,i) = \pm e_i}$ for ${i = 1, \ldots, p_{11}}$ and hence $U_{11} = J_{11}$ is a signed identity matrix.

\section{Proof of Lemma \ref{S2lem}} \label{app:pf_s2}

\begin{IEEEproof} The proof is given for the case where $p_3 = p_2 = 0$, for which the notation is considerably simpler. The proof of the general case follows in exactly the same manner. In this case, $A_{12}$, $A_{22}$, $B_1$ and $B_2$ are given by:
\beq \label{Vnmp4}
\left[ \barr {c|c} A_{12} & B_1 \\ \hline A_{22} & B_2 \earr \right] = 
\left[ \barr {ccc|c}
0 & I & 0 & 		0\\
\hline
\times & \times & \al_{14} & 			 I \\
\al_{31} & \times & \al_{34} &			 0\\
\times & \times & \al_{44} & 			 0\earr \right]
\eeq

\noindent Equations \eqref{gl3} and \eqref{gl4} now simplify to:
\begin{subequations} \label{gl34s}
\begin{align}
&A_{12}S_2 = 0 \label{A12S2pf}\\
&S_2A_{22}^T + A_{22}S_2 +B_2B_2^T - T_2B_2B_2^TT_2^T = 0 \label{lyap}
\end{align}
\end{subequations}

\noindent from which $S_2$ and $T_2B_2$ are required to be in the following forms, partitioned as $A_{22}$ and $B_2$:
\beq \label{S2}
S_2 = \bmat s_{11} & 0 & s_{12}  \\  0 & 0 & 0 \\ s_{12}^T & 0 & s_{22} \emat \qquad \textrm{and} \qquad
T_2B_2 = \bmat t_{1} \\ 0 \\ t_{3} \emat
\eeq

\noindent We will now prove by induction that we must have $s_{11} = 0$ and $s_{12}=0$ to satisfy \eqref{gl34s} for any valid choice of $T_2$. 

Recall that for $i = 1, \ldots, p_1$, the number $k_i$  is the smallest value of $j$ in the range $0 < j < l$ such that ${\al^{(j)}_{31}(i,i) \neq 0}$. Hypothesize that the following statement holds for ${i = 1, \ldots, p_1}$ and for all $j = 1, \ldots, k_i-1$:

\beq \label{H}
\boxed{
\bali
\al^{(j)}_{34}(i,:)s_{12}^T &= 0^T\\
\al^{(j)}_{34}(i,:)s_{22} &= 0^T
\eali
}
\eeq

\subsection*{Base case: ${j=1}$}

\noindent Multiply \eqref{lyap} by $A_{12}(i,:)$ on the left for some $i$ in the range $1 \leq i \leq p_1$ with $k_i > 1$:
\beqs
\bali
&A_{12}(i,:)S_2A_{22}^T + A_{12}(i,:)A_{22}S_2\\
&+ A_{12}(i,:)B_2B_2^T - A_{12}(i,:)T_2B_2B_2^TT_2^T = 0^T
\eali
\eeqs

\noindent and note that directly from \eqref{Vnmp4} and \eqref{S2} terms one, three and four are zero. Hence we have:
\beq \label{k1H1}
A_{12}(i,:)A_{22}S_2 = 0^T
\eeq

\noindent Since $\al_{31}(i,:) =0^T$ (because $k_i > 1$), \eqref{k1H1} gives:
\beqs \label{k1H2}
\al_{34}(i,:)s_{12}^T = 0^T \qquad \mathrm{and} \qquad \al_{34}(i,:)s_{22} = 0^T
\eeqs

\noindent The hypothesis \eqref{H} therefore holds for $j=1$ for all $i=1, \ldots, p_1$ with $k_i >1$.

\subsection*{Induction}

\noindent Suppose for some $i$ in the range $1 \leq i \leq p_1$, for some $\hk < k_i$, the hypothesis holds for $j=\hk-1$. This implies that:
\beq \label{k_H2}
\al^{(\hk-1)}_{34}(i,:)s_{12}^T = 0^T \quad \mathrm{and} \quad \al^{(\hk-1)}_{34}(i,:)s_{22} = 0^T
\eeq

\noindent Now show that the hypothesis is satisfied for $j=\hk$ as follows. First multiply \eqref{lyap} on the left by $A_{12}(i,:)A_{22}^{\hk-1}$ to give:
\beq \label{klyap}
\bali
&A_{12}(i,:)A_{22}^{\hk-1}S_2A_{22}^T + A_{12}(i,:)A_{22}^{\hk}S_2\\
&+ A_{12}(i,:)A_{22}^{\hk-1}B_2B_2^T\\
&- A_{12}(i,:)A_{22}^{\hk-1}T_2B_2B_2^TT_2^T = 0^T
\eali
\eeq

\noindent Since $\al^{(\hk-1)}_{31}(i,:) = 0^T$, the expression $A_{12}(i,:)A_{22}^{\hk-1}S_2$ is equal to zero from \eqref{k_H2} and hence the first term in \eqref{klyap} is equal to zero. The third term is also zero due to ${\al^{(\hk-1)}_{31}(i,:) = 0^T}$.  The remaining two terms give:
\begin{subequations} \label{ki1_2}
\begin{align}
\al^{(\hk)}_{34}(i,:)s_{12}^T - \al^{(\hk-1)}_{34}(i,:) t_{3}t_{1}^T  &= 0^T \label{ki1} \\
\al^{(\hk)}_{34}(i,:)s_{22}^T - \al^{(\hk-1)}_{34}(i,:) t_{3}t_{3}^T  &= 0^T \label{ki2} 
\end{align}
\end{subequations}

\noindent where we have used the fact that $\al^{(\hk)}_{31}(i,:) = 0^T$ since ${k_i > \hk}$. Now multiply \eqref{ki2} on the right by $\al^{(\hk-1)}_{34}(i,:)^T$, which, from \eqref{k_H2}, gives:
\beqs
\al^{(\hk-1)}_{34}(i,:) t_{3}t_{3}^T  \al^{(\hk-1)}_{34}(i,:)^T = 0
\eeqs

\noindent which implies $\al^{(\hk-1)}_{34}(i,:)t_3 = 0^T$. This eliminates all $T_2$ terms from \eqref{ki1_2}, giving the desired result:
\beqs \label{ki1_2s}
\al^{(\hk)}_{34}(i,:)s_{12}^T  = 0^T \qquad \mathrm{and} \qquad \al^{(\hk)}_{34}(i,:)s_{22}^T = 0^T
\eeqs

\noindent By induction the hypothesis \eqref{H} therefore holds for all $i = 1, \ldots, p_1$ for $j = 1, \ldots, k_i-1$.

\subsection*{Termination}

\noindent To show that $s_{11}$ and $s_{12}$ must be equal to zero, multiply \eqref{lyap} on the left by $A_{12}(i,:)A_{22}^{k_i-1}$ for any $i$ such that $1 \leq i \leq p_1$. Recall that:
\beqs
\bali
\al^{(k_i)}_{31}(i,:) &= \al^{(k_i)}_{31}(i,i) e_i^T \neq 0\\
\al^{(k_i)}_{34}(i,:) &= 0^T
\eali
\eeqs

\noindent and hence the equivalent of \eqref{ki1_2} is:
\begin{subequations} \label{ki1_2T}
\begin{align}
\al^{(k_i)}_{31}(i,i)s_{11}(i,:) - \al^{(k_i-1)}_{34}(i,:) t_{3}t_{1}^T &= 0^T \label{ki1T} \\
\al^{(k_i)}_{31}(i,i)s_{12}(i,:) - \al^{(k_i-1)}_{34}(i,:) t_{3}t_{3}^T &= 0^T \label{ki2T} 
\end{align}
\end{subequations}

\noindent Since the hypothesis \eqref{H} holds for $j=k_i-1$, we know that $\al^{(k_i-1)}_{34}(i,:)s_{12}^T  = 0^T$. Multiply \eqref{ki2T} on the right by $\al^{(k_i-1)}_{34}(i,:)^T$ to give $\al^{(k_i-1)}_{34}(i,:) t_{3} = 0^T$ and \eqref{ki1_2T} then simplifies to:
\beqs
\bali
\al^{(k_i)}_{31}(i,i)s_{11}(i,:)  &= 0^T \qquad \Rightarrow \qquad s_{11}(i,:)  = 0^T\\
\al^{(k_i)}_{31}(i,i)s_{12}(i,:)  &= 0^T \qquad \Rightarrow \qquad s_{12}(i,:)  = 0^T
\eali
\eeqs

\noindent Since the above holds for every $i = 1, \ldots, p_1$, we therefore have $s_{11} = 0$ and $s_{12} = 0$.
\end{IEEEproof}

%
%
%
%
\newpage

\bibliographystyle{./IEEEtran}
\bibliography{./noise}

\end{document}